\documentclass[conference]{IEEEtran}
\IEEEoverridecommandlockouts
\usepackage{cite,graphicx,amssymb,amsmath,color,textcomp,array,amsthm,tabulary,tabularx,mathtools}
\usepackage{extarrows,multirow,multicol,enumitem }
\usepackage{bm}
\usepackage{subeqnarray}
\usepackage{subfig}
\usepackage{algorithm}
\usepackage{algpseudocode}
\usepackage{soul}

\usepackage{booktabs}

\newtheorem{remark}{\underline{Remark}}

\newlength{\figwidth}
\begin{document}

\title{ 
Multi-Shot Quantum Sensing for RF Signal Detection with MIMO Rydberg-Atom Receivers
\vspace{-3mm}
}

\author{
    \IEEEauthorblockN{Saman Atapattu\IEEEauthorrefmark{1}, 
    Harini Hapuarachchi\IEEEauthorrefmark{2}, and 
    Nathan Ross\IEEEauthorrefmark{3}}
    \IEEEauthorblockA{
        \IEEEauthorrefmark{1}Department of Electrical and Electronic Engineering, 
 School of Engineering, 
RMIT University, Melbourne, 
Australia\\
        \IEEEauthorrefmark{2}School of Electrical and Computer Engineering, University of Sydney, Sydney, Australia\\
        \IEEEauthorrefmark{3}School of Mathematics and Statistics, University of Melbourne, Melbourne, Australia\\
        Email: 
\IEEEauthorrefmark{1}saman.atapattu@rmit.edu.au,
\IEEEauthorrefmark{2}harini.hapuarachchi@sydney.edu.au,
\IEEEauthorrefmark{3}nathan.ross@unimelb.edu.au
    }
\vspace{-10mm}}

\maketitle

\begin{abstract}
Rydberg-atom quantum receivers (RAQRs) enable electric-field sensing with quantum-noise–limited performance, yet their optical readout provides only magnitude measurements whose fluctuations follow Rician statistics governed by atomic projection noise, optical shot noise, reference-field injection, and short coherence times. These non-Gaussian, phase-blind measurements invalidate classical single-shot RF detectors and necessitate multi-shot
quantum sensing strategies. This work develops a physically consistent multi-shot statistical model for RAQRs and derives both the optimal genie-aided likelihood-ratio test (LRT) and a practical phase-averaged LRT that removes dependence on the unknown RF-field phase. Closed-form test statistics and thresholds are obtained for both detectors, and the limits imposed by finite quantum shots—due to atomic dephasing and measurement backaction—are explicitly quantified. A fully non-coherent energy detector is also analysed, with exact detection probability derived using noncentral chi-square models. Monte Carlo results show that only $5$--$10$ quantum shots yield major gains: the phase-averaged LRT closely approaches the genie bound and RAQR detection markedly outperforms classical RF energy detection under comparable received power. The proposed framework provides the first unified statistical basis for multi-shot Rydberg-based weak-field detection and underscores the potential of RAQRs for quantum-enhanced signal detection.
\end{abstract}

\vspace{-0mm}

\section{Introduction}

Quantum RF sensors based on Rydberg atoms have emerged as a powerful platform
for detecting ultra-weak electromagnetic fields, in some regimes surpassing
classical receiver sensitivity~\cite{Li2025awpl,Wan2025awpl}. Unlike
conventional front ends that rely on mixers and coherent downconversion, a
Rydberg-atom quantum receiver (RAQR) directly maps incident RF fields onto
optical transmission via electromagnetically induced transparency (EIT),
enabling room-temperature, quantum-noise-limited readout~\cite{Cui2025magc,
Gong2025qcnc}. This quantum transduction enables femtowatt-level sensing and
supports applications in covert communication, spectrum awareness, and
non-invasive RF monitoring~\cite{Guo2025tvt,Atapattu2025comll}.
RAQR measurements differ fundamentally from classical I/Q samples: each
interrogation yields a single \emph{magnitude-only} value whose fluctuations
follow a Rician law~\cite{Rice1948bell} determined by atomic projection noise, optical shot noise,
and the injected reference field~\cite{Cui2025jsac,Guo2025wcl}. This
non-Gaussian structure limits the effectiveness of classical single-shot
detectors and motivates principled \emph{multi-shot quantum sensing}
frameworks that explicitly incorporate Rydberg readout statistics into
detection theory~\cite{Atapattu2025comll,Cui2025jsac}.

\vspace{-2mm}
\subsection{Related Work}
\label{sec:related_work}
\vspace{-1mm}
Recent literature underscores the potential of RAQRs as a credible alternative to classical RF systems. Foundational surveys outline their principles and advantages in sensitivity and form factor \cite{Cui2025magc}, while subsequent works have developed more accurate end-to-end signal models and digital linearization techniques to manage their inherent nonlinearity \cite{Gong2025qcnc,Xia2024wcsp}. A significant research thrust addresses the core challenge of ``magnitude-only'' detection. For multiple-input-multiple-output (MIMO) communications, this has led to capacity-achieving precoding schemes \cite{Cui2025icc} and specialized channel estimation algorithms \cite{Xu2025wcl}. In wireless sensing, phase-retrieval-based methods have been adapted for angle-of-arrival (AoA) estimation and multi-user detection \cite{Cui2025jsac,Kim2025comml}, with a notable demonstration of single-sensor AoA recovery via intra-vapor interference \cite{Guo2025tcom}.
From a detection-theoretic perspective, \cite{Atapattu2025comll} formulated constant false alarm rate (CFAR) detectors and likelihood-ratio tests (LRTs) for single-shot measurements, and an information-theoretic analysis confirmed that the RAQR's immunity to thermal noise can offset the capacity loss from its phase-blind nature \cite{Guo2025wcl}. Parallel hardware innovations, such as high-sensitivity enhancement resonators \cite{Li2025awpl,Wan2025awpl} and broadband magnetic tuning architectures \cite{Qimeng2025sj}, continue to push performance boundaries. 

\vspace{-1mm}
\subsection{Contributions}
\vspace{-1mm}
While existing work has established models and single-shot detection methods for RAQRs, a statistical theory for \emph{multi-shot} detection—which aggregates several quantum measurements to improve reliability—remains unexplored. Current approaches, even those based on LRTs, typically rely on single-shot thresholding \cite{Atapattu2025comll}, leaving the full sensing potential of RAQRs untapped. \textit{This paper bridges that gap by developing a comprehensive detection framework for multi-shot RAQR measurements}. Our specific contributions are:

\begin{figure*}[htp]
    \centering
    \includegraphics[width=\textwidth]{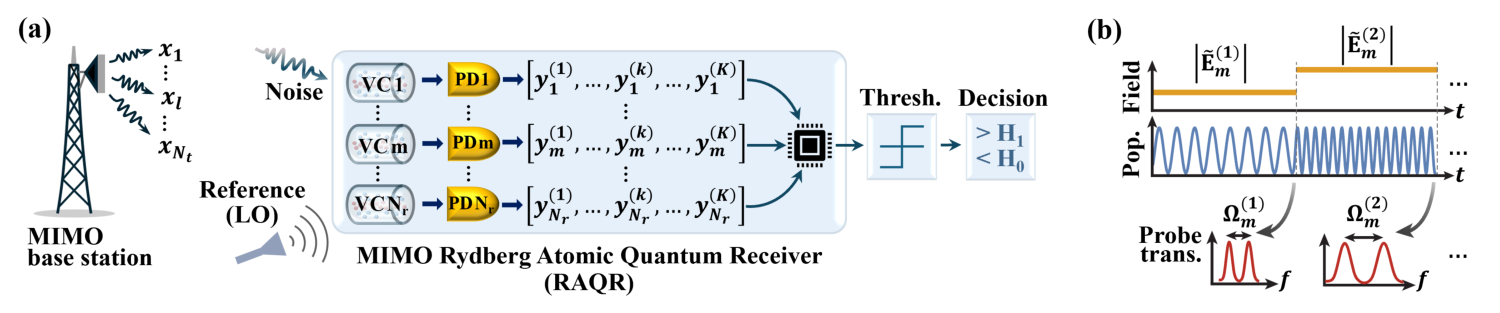}
    \caption{(a) RF--quantum MIMO architecture in which a
    RAQR performs multi-shot optical readout.
    (b) Conceptual illustration of consecutive interrogation shots. For
    clarity, decoherence-induced damping of population oscillations is shown
    as negligible.\vspace{-5mm}}
    \label{f_Rydberg}
\end{figure*}

\begin{itemize}
    \item \textit{Exact Multi-Shot LRT:} We derive the first multi-shot likelihood-ratio test for RAQR magnitude measurements, formalizing the underlying product-Rician statistical structure that arises from atomic noise and reference.

    \item \textit{Practical Phase-Averaged LRT:} For scenarios with unknown signal phases (e.g., phase-shift keying (PSK)), we develop an implementable detector that marginalizes over the phase uncertainty using Rician moments, resulting in an LRT that requires no explicit waveform knowledge.

    \item \textit{Non-Coherent Quantum Energy Detector (ED):} We analyze aa  ED suitable when neither the channel nor the signal is known, and derive closed-form CFAR thresholds based on noncentral chi-square statistics.

    \item \textit{Performance Analysis of Quantum Multi-Shot Gain:} Through numerical simulations, we demonstrate that aggregating a small number of shots ($K\!=\!3$--$8$) yields significant performance gains. The proposed phase-averaged LRT recovers most of the performance of a genie-aided detector and substantially outperforms classical RF energy detectors operating at standard thermal noise levels.
\end{itemize}

This work establishes the first statistical framework for multi-shot quantum RF detection, highlighting its potential for quantum-enhanced spectrum sensing, weak-signal detection, and hybrid quantum–classical RF systems~\cite{Obando2025vtc}.

\section{System Overview}\label{sec:sys}

We consider a hybrid RF--quantum sensing scenario in which a multi-antenna
primary transmitter communicates with its users while a secondary node
monitors the spectrum using one or more Rydberg-atom quantum receivers
(RAQRs). As shown in Fig.~\ref{f_Rydberg}(a), the RAQR replaces the
conventional RF front end and forms the first stage of a MIMO
RF--quantum sensing architecture.

\subsubsection{Quantum Sensing Model}
A RAQR operates by mapping the incident RF field onto the optical response of
a Rydberg atomic ensemble. Each interrogation produces a
\emph{magnitude-only} optical readout (Fig.~\ref{f_Rydberg}(b)), whose
fluctuations are dominated by atomic projection noise, photon shot noise, and
the injected reference field rather than electronic thermal noise.  
Because each probe collapses the atomic state, only a limited number of
measurements---typically a few to a few tens within the atomic coherence
time---can be collected per sensing interval. These $K$ magnitude samples from
$N_r$ vapor cells form the input to the multi-shot detectors developed here.

\subsubsection{Binary Detection Task}
The sensing objective is to determine whether an unknown RF emitter is
present:
\begin{align*}
    \mathcal{H}_0:\quad& \text{reference field + quantum noise only},\\
\mathcal{H}_1:\quad& \text{additional unknown RF field present}.
\end{align*}
Because the RAQR provides only noisy magnitudes, reliable discrimination
requires aggregating multiple shots.

\subsubsection{Network Interpretation}
This model captures a range of wireless tasks: (i) spectrum sensing in
cognitive radio; (ii) weak-emitter or unknown-signal detection; (iii)
jammer or covert-transmitter identification; and (iv) interference monitoring
in dense networks. In all cases, the RAQR acts as a quantum-enhanced
front end that must infer emitter activity from a small set of
non-Gaussian magnitude measurements.

\subsubsection{Role of Multi-Shot Processing}
Single-shot RAQR readouts are strongly noise-limited.  
Accumulating multiple interrogations stabilizes the underlying Rician
statistics and enables reliable hypothesis testing.  
The detectors developed in this paper explicitly quantify performance as a
function of the shot budget~$K$ and provide quantum-aware strategies that
remain effective under realistic coherence-time constraints.

\vspace{-2mm}
\section{Analytical Models}

Our objective is to develop a system and analytical framework that captures  
(i) the classical MIMO propagation from the primary base station (BS) and  
(ii) the quantum–optical measurement process inherent to the Rydberg-atom receiver.  
The remainder of this section establishes the models required for this joint RF–quantum sensing problem.

\vspace{-2mm}
\subsection{Network Model}

We consider a primary MIMO transmitter with \(N_t\) co-located antennas and a
secondary spectrum-sensing receiver equipped with \(N_r\) Rydberg-atom vapor
cells (VCs), each coupled to an optical probe and photodetector, as shown in
Fig.~\ref{f_Rydberg}.  
A weak RF reference field, derived from a local oscillator (LO), is injected at
each VC to stabilise the atomic response during measurement.
The BS transmits an \emph{unknown deterministic} constant-modulus vector
\begin{equation}
    \mathbf{x} = [x_1,\dots,x_{N_t}]^\mathsf{T}, 
    \qquad x_\ell = \sqrt{P_\ell}\, s_\ell,
    \label{eq:x_signal}
\end{equation}
where \(s_\ell\) is a unit-modulus baseband symbol and \(P_\ell\) is the
per-antenna transmit power.  
The total transmit power satisfies
\(
    \|\mathbf{x}\|^2 = \sum_{\ell=1}^{N_t} |x_\ell|^2 = P ,
\)
and under constant-modulus signalling (e.g., \(M\)-PSK) this reduces to
\(
    \sum_{\ell=1}^{N_t} P_\ell = P,
\)
with equal-power allocation \(P_\ell = P/N_t\) commonly assumed.

The antennas operate at a carrier frequency resonant with the RF-driven
Rydberg transition, and the radiated field propagates through a frequency-flat,
quasistatic MIMO channel to the \(N_r\) VCs.  
Each VC performs \(K\) interrogation cycles—constituting a \emph{multi-shot}
measurement process (in contrast to the single-shot model of~\cite{Atapattu2025comll})—and
produces \(K\) optical readouts that are jointly processed for spectrum sensing.

The total sensing duration \(T_{\mathrm{sig}} = K T_s\) is selected to lie within
the channel coherence time, ensuring that the BS-to-VC channel remains
approximately constant across all \(K\) shots.

\subsection{Atom--Field Quantum Interaction}

Each vapor cell (VC) contains an ensemble of alkali atoms prepared in a
Rydberg configuration using two counter–propagating lasers: a weak probe beam
driving the \(|1\rangle \!\rightarrow\! |2\rangle\) transition and a strong coupling
beam driving \(|2\rangle \!\rightarrow\! |3\rangle\) \cite{Cui2025jsac,Gong2025IEEEWireless}.
This establishes a ladder-type electromagnetically induced transparency (EIT)
system whose upper state \(|3\rangle\) is highly sensitive to incident RF fields.
An RF field resonant with the \(|3\rangle \leftrightarrow |4\rangle\) transition
modifies the optical steady state through Autler--Townes (AT) splitting, and the
corresponding change in probe transmission is recorded by the photodetector,
yielding an estimate of the effective RF-driven Rabi frequency magnitude
\(\Omega_m^{(k)}\) during the \(k\)-th interrogation shot of VC~\(m\)
\cite{Cui2025jsac,fancher2021TQE}.
The complex envelope of the RF field received at VC \(m\) during shot \(k\) is
\begin{equation}
    \tilde{\mathbf{E}}_m^{(k)}
    = \frac{1}{2}\!\left(
        \sum_{\ell=1}^{N_t}
        \boldsymbol{\hat{\epsilon}}_{m,\ell}\,
        x_\ell \sqrt{\rho_{m,\ell}}\, g_{m,\ell}
        + \mathbf{R}_m
        + \mathbf{W}_{m,E}^{(k)}
      \right),
\end{equation}
where \(\rho_{m,\ell}\) denotes large-scale attenuation (assumed equal across
\(\ell,m\) due to the co-located geometry), \(g_{m,\ell}\in\mathbb{C}\) models
small-scale fading and RF phase, and
\(\boldsymbol{\hat{\epsilon}}_{m,\ell}\) is the polarization unit vector seen by
VC \(m\).  
The term \(\mathbf{R}_m\) is the received LO-derived reference field used to
linearise and stabilise the atomic response, while
\(\mathbf{W}_{m,E}^{(k)}\) represents extrinsic free-space noise during shot
\(k\).  
Although the transmitted waveform and reference are constant across shots,
each VC experiences a distinct spatially filtered version of the field due to its
location, channel coefficients, and polarization coupling.

At the quantum level, the received field determines the magnitude of the
Rabi coupling \(\Omega_m^{(k)}\propto
|\boldsymbol{\mu}_{34}^\mathsf{H}\tilde{\mathbf{E}}_m^{(k)}|\), where
\(\boldsymbol{\mu}_{34}\) is the transition dipole moment.  
This Rabi frequency modulates the optical transparency, and the photodetected
AT response produces the magnitude-only measurement used in our statistical
model.

\subsection{Multi-Shot Measurement}

We assume that consecutive atomic interrogations are temporally separated
such that the noise samples across shots are effectively uncorrelated.
Each shot is taken only after the RF-driven Rabi-frequency magnitude
\(\Omega_m^{(k)}\) 
has reached a
quasi–steady state within the shot duration \(T_s\), as illustrated in
Fig.~\ref{f_Rydberg}(b).  
Under these constraints, the number of usable and approximately
independent measurements within a signalling interval \(T_{\mathrm{sig}}\) is
limited by
\begin{equation}
    K \;\lesssim\; \left\lfloor
        \frac{T_{\mathrm{sig}}}{\max(T_R,\,T_N)}
    \right\rfloor,
\end{equation}
where \(T_R\) denotes the AT-splitting (Rabi-magnitude) settling time and
\(T_N\) denotes the noise correlation time determined by atomic and
photon-shot fluctuations.

Experimental studies of room-temperature multi-photon Rydberg excitation
show transient EIT/AT responses settling within a few microseconds
\cite{bohaichuk2023three}.  
The noise correlation time \(T_N\) is typically comparable to the atomic
coherence time, which ranges from hundreds of nanoseconds to a few
microseconds depending on the Rydberg level, vapor-cell temperature, and
optical-beam configuration \cite{adams2020rydberg}.  
Thus, practical operating conditions often yield
\(T_R,\,T_N \sim 1\text{--}10~\mu\mathrm{s}\).

\emph{Example:}  
In a favourable room-temperature configuration with
\(T_N \sim 1\text{--}5~\mu\mathrm{s}\) and \(T_R \sim 2\text{--}10~\mu\mathrm{s}\),
a signalling block aligned with a typical wireless channel coherence
interval of \(T_c \approx 125~\mu\mathrm{s}\) supports approximately
\(
K \;\approx\; \left\lfloor
    \frac{T_c}{\max(T_R,T_N)}
\right\rfloor \;\approx\; 10\text{--}12
\)
effectively independent measurement shots per vapor cell.
These multi-shot observations form the basis of the statistical detection
framework developed in this work.

\subsection{Rabi-Frequency Mapping of the Received Field}

Following the steady-state Rydberg–EIT/AT formalism in
\cite{Cui2025jsac,Kim2025comml}, the stabilised atom–field coupling strength
(i.e., Rabi-frequency magnitude) measured at the \(m\)-th vapor cell during the
\(k\)-th interrogation is
\begin{equation}
    \Omega_m^{(k)}
    = \left|
        \sum_{\ell=1}^{N_t} h_{m,\ell}\, x_\ell
        + r_m
        + w_{m,E}^{(k)}
      \right|,
    \label{eq:Rabi_map}
\end{equation}
where the effective complex channel gain  
\(
    h_{m,\ell}
    = \frac{1}{\hbar}\,
      \boldsymbol{\mu}_{eg}^H
      \boldsymbol{\hat{\epsilon}}_{m,\ell}\,
      \sqrt{\rho_{m,\ell}}\,
      g_{m,\ell}
\)
captures the atomic dipole projection, RF-field polarization alignment, path
loss, and small-scale fading, all assumed quasistatic over the signalling
interval. The term
\(r_m = \boldsymbol{\mu}_{eg}^H\mathbf{R}_m/\hbar\)
is the received component of the common LO reference field, while  
\(w_{m,E}^{(k)}=\boldsymbol{\mu}_{eg}^H\mathbf{W}_{m,E}^{(k)}/\hbar
 \sim\mathcal{CN}(0,\sigma_{E,m}^2)\)
denotes the projected extrinsic RF noise during the \(k\)-th shot.
Under the deterministic signalling model introduced earlier, the transmitted
symbols satisfy \(x_\ell^{(k)} = x_\ell\) for all \(k\), so only the projected
noise term varies across shots. The mapping in \eqref{eq:Rabi_map} therefore
provides a direct magnitude-only measurement of a coherent RF superposition
plus noise and reference injection, as depicted in Fig.~\ref{f_Rydberg}(b).

Both continuous-wave (CW) and pulsed multi-shot operation are compatible with
\eqref{eq:Rabi_map}. In a CW scheme, the probe and coupling beams remain on
throughout the signalling interval, and the photodetector samples the AT
splitting after the Rabi magnitude has stabilised within each shot duration
\(T_s\). In pulsed operation, the optical beams are re-applied for each shot,
but the same steady-state expression applies provided that AT-splitting
stabilisation is reached within \(T_s\). In both cases, the statistical model
of \(\Omega_m^{(k)}\) is identical, differing only in the physical
implementation of the measurement cycle.

\subsection{Atomic Receiver Signal Model}\label{ss_AR_sys}

Each Rydberg vapor cell produces a magnitude-only readout obtained from the
steady-state Rabi frequency extracted via the probe transmission
\cite{Cui2025jsac}. The measurement at the \(m\)-th cell during the \(k\)-th
shot (Fig.~\ref{f_Rydberg}) is modeled as
\begin{align}
    y_m^{(k)}
    \!=\! \Big| \sum_{\ell=1}^{N_t} h_{m,\ell} x_\ell \!+\! r_m \!+\! w_m^{(k)} \Big|
       \!=\! \big| [\mathbf{H}\mathbf{x}]_m \!+\! r_m \!+\! w_m^{(k)} \big|
    \label{eq:ymk_scalar}
\end{align}
where \([\mathbf{H}\mathbf{x}]_m\) is the \(m\)-th element of the received
composite field. The full MIMO channel matrix is $\mathbf{H} = [h_{ij}] \in \mathbb{C}^{N_r \times N_t}$, 
where $h_{ij}$ denotes the channel coefficient from transmit antenna $j$ to receive antenna $i$.
The transmit vector \(\mathbf{x}\) follows the unknown deterministic
constant-modulus model in~\eqref{eq:x_signal}. The reference term \(r_m\) is the
received component of the injected LO field. 
Stacking the noise across all shots yields
\begin{align}
    \mathbf{W}=[\mathbf{w}^{(1)},\dots,\mathbf{w}^{(K)}], \qquad
    \mathbf{w}^{(k)}\sim\mathcal{CN}(\mathbf{0},\sigma_w^2\mathbf{I}_{N_r}),
\end{align}
with entries
\(
[\mathbf{W}]_{m,k}=w_m^{(k)}=w_{m,E}^{(k)}+w_{m,I}^{(k)}.
\)
Here \(w_{m,E}^{(k)}\) is projected RF/environmental noise and \(w_{m,I}^{(k)}\)
collects internal noise sources such as quantum projection noise, photon-shot
noise, and electronics. Consistent with prior
RAQR-based sensing models
\cite{Kim2025coml,Gong2025IEEEWireless,Atapattu2025comll},
we adopt the tractable and experimentally supported CSCG approximation
\(
w_m^{(k)}\sim \mathcal{CN}(0,\sigma_w^2)\), where \(
\sigma_w^2=\sigma_{E,m}^2+\sigma_{I,m}^2,
\)
with samples assumed i.i.d.\ across receivers and shots. Under this model, each
RAQR output is a Rician-distributed magnitude of a coherent RF field plus
quantum-dominated noise, greatly differing from classical coherent I/Q samples
and forming the basis of the detection framework developed next.

\subsection{Detection Problem Formulation}

We consider a \emph{multi-shot} sensing model in which each of the \(N_r\)
Rydberg receivers produces \(K\) temporally independent magnitude
measurements. As illustrated in Fig.~\ref{f_Rydberg}(a), collecting all observations gives
\[
\mathbf{Y}=
\begin{bmatrix}
y_1^{(1)} & \cdots & y_1^{(K)}\\[-1mm]
\vdots    & \ddots & \vdots    \\[-1mm]
y_{N_r}^{(1)} & \cdots & y_{N_r}^{(K)}
\end{bmatrix}
\in\mathbb{R}_+^{N_r\times K}.
\]

The goal is to determine whether an unknown RF field is \emph{absent} or
\emph{present}. Under the magnitude-only RAQR model~\eqref{eq:ymk_scalar}, the
binary hypothesis test becomes
\begin{align}
\mathbf{Y} =
\begin{cases}
|\mathbf{R} + \mathbf{W}|, 
& \mathcal{H}_0:\ \text{no RF signal},\\[1mm]
|\mathbf{H}\mathbf{X} + \mathbf{R} + \mathbf{W}|, 
& \mathcal{H}_1:\ \text{RF signal present},
\end{cases}
\label{eq_H0H1_multi}
\end{align}
where the block-structured quantities are
\[
\mathbf{X}=\mathbf{x}\mathbf{1}_K^T,\qquad
\mathbf{R}=\mathbf{r}\mathbf{1}_K^T,\qquad
\mathbf{W}\!=\![\mathbf{w}^{(1)},\ldots,\mathbf{w}^{(K)}],
\]
and \(|\cdot|\) denotes elementwise magnitude. The unknown deterministic vector
\(\mathbf{x}\in\mathbb{C}^{N_t}\) represents the BS transmission during the
coherence interval; \(\mathbf{r}\in\mathbb{C}^{N_r}\) is the received reference
field; and \(\mathbf{H}\in\mathbb{C}^{N_r\times N_t}\) is the quasistatic MIMO
channel mapping the primary transmitter to the Rydberg sensors.

\paragraph*{Temporal assumptions.}
Throughout a wireless coherence block, the tuple \((\mathbf{H},\mathbf{x},
\mathbf{r})\) remains constant. Shot-to-shot randomness originates solely from
independent atomic and optical noise samples
\(
[\mathbf{W}]_{m,k}=w_m^{(k)}\sim\mathcal{CN}(0,\sigma_w^2)\), for \(m=1\!:\!N_r,\ k=1\!:\!K,
\)
which are assumed i.i.d.\ across cells and shots. This model captures the
physical constraints of RAQR operation: deterministic incident fields within a
coherence interval, but independent quantum-limited fluctuations across
interrogation shots.
This multi-shot magnitude framework forms the basis for the likelihood-ratio
derivations and detector designs presented in the following sections.
\section{Multi-Shot Likelihood Ratio Test (LRT)}
\label{sec:lrt_multishot}

Quantum sensors such as Rydberg-atom receivers do not produce coherent
I/Q samples; instead, each interrogation yields a noisy \emph{magnitude-only}
measurement whose statistics follow a Rician law shaped by atomic quantum
noise, reference-field injection, and the shot-to-shot reset of the atomic
ensemble. As a result, classical RF detection techniques---which rely on
Gaussian baseband models---do not apply directly.  
This section establishes the likelihood functions associated with the
multi-shot RAQR model and derives the corresponding LRTs that form the
statistical foundation for quantum-enhanced RF detection.

From the multi-shot model in~\eqref{eq_H0H1_multi}, each observation satisfies
\begin{align}
y_m^{(k)} =
\begin{cases}
|r_m + w_m^{(k)}|, 
& \mathcal{H}_0,\\[1mm]
|\mathbf{h}_m^T\mathbf{x} + r_m + w_m^{(k)}|, 
& \mathcal{H}_1,
\end{cases}
\quad k=1,\ldots,K ,
\label{eq:multishot_model}
\end{align}
where the noise samples \(w_m^{(k)}\sim\mathcal{CN}(0,\sigma_w^2)\) are
independent across vapor cells \(m\) and shots \(k\).

\subsection{Likelihood Functions}

Under either hypothesis, the complex quantity inside the magnitude in
\eqref{eq:multishot_model} is Gaussian with mean \(\mu_{m,0}=r_m\) under
\(\mathcal{H}_0\) and \(\mu_{m,1}=\alpha_m(\mathbf{x})\triangleq
\mathbf{h}_m^T\mathbf{x}+r_m\) under \(\mathcal{H}_1\).  
Therefore each \(y_m^{(k)}\) is Rician-distributed~\cite{Rice1948bell}, and independence across
\((m,k)\) yields factored likelihoods. 

Under $\mathcal{H}_0$,
\begin{align}
p(\mathbf{Y}|\mathcal{H}_0)
\!=\!\!
\prod_{m=1}^{N_r}\prod_{k=1}^{K}
\frac{y_m^{(k)}}{\sigma_w^2}
I_0\!\left(\!
\frac{2y_m^{(k)}|r_m|}{\sigma_w^2}\!
\right)
e^{
-\frac{(y_m^{(k)})^2 + |r_m|^2}{\sigma_w^2}},
\label{eq:like_H0}
\end{align}
where \( I_0(\cdot) \) is modified Bessel function of the first kind~\cite{gradshteyn2014table}.

Under $\mathcal{H}_1$,
\begin{align}
p(\mathbf{Y}|\mathcal{H}_1,\mathbf{x})
=
\prod_{m=1}^{N_r}\prod_{k=1}^{K}&
\frac{y_m^{(k)}}{\sigma_w^2}
I_0\!\left(
\frac{2y_m^{(k)}|\alpha_m(\mathbf{x})|}{\sigma_w^2}
\right)
\notag\\[-1mm]
&\qquad\times
e^{
-\frac{(y_m^{(k)})^2 + |\alpha_m(\mathbf{x})|^2}{\sigma_w^2}
}.
\label{eq:like_H1}
\end{align}


The likelihood-ratio test (LRT) is defined as~\cite{trees2001detection}
\begin{align}
\Lambda(\mathbf{Y})
= \frac{p(\mathbf{Y}|\mathcal{H}_1)}{p(\mathbf{Y}|\mathcal{H}_0)}
\underset{\mathcal{H}_0}{\overset{\mathcal{H}_1}{\gtrless}}
\eta,
\label{eq_LRT_def}
\end{align}
where \(\eta\) is the MAP threshold determined by the prior probability
\(P(\mathcal{H}_1)\). 

A central challenge is that the BS transmit vector $\mathbf{x}$ is unknown to
the RAQR. This motivates two detector classes:

\begin{enumerate}
    \item \textit{Genie-aided LRT:}  
          assumes perfect knowledge of $\mathbf{x}$.  
          This detector is not implementable but provides a fundamental
          quantum-sensing performance bound.

    \item \textit{Phase-averaged LRT (practical):}  
          exploits the constant-modulus structure of $M$-PSK signalling and
          replaces $|\alpha_m(\mathbf{x})|$ with its statistical expectation,
          yielding a computationally efficient detector. 
\end{enumerate}

These two LRTs provide (i) a quantum-limited benchmark and (ii) an implementable
RAQR-specific detection strategy that respects the magnitude-only 
Rician measurement.

\subsection{Genie-Aided LRT}

For fixed $(\mathbf{H},\mathbf{x},\mathbf{r})$, each RAQR observation in
\eqref{eq:multishot_model} is Rician with noncentrality 
$|r_m|$ under $\mathcal{H}_0$ and $|\alpha_m|$,  
$\alpha_m \!=\! \mathbf{h}_m^T\mathbf{x}+r_m$, under $\mathcal{H}_1$.
Using \eqref{eq:like_H0}–\eqref{eq:like_H1}, cancelling common factors,
and taking logarithms gives the \emph{exact} multi-shot log-likelihood ratio
\begin{align}
\ln\Lambda(\mathbf{Y})
\!\!=\!\!
\sum_{m=1}^{N_r}\sum_{k=1}^{K}
\Biggl[
\frac{|r_m|^2 \!-\! |\alpha_m|^2}{\sigma_w^2}
\!+\! \ln\!
\frac{
I_0\!\left(\tfrac{2y_m^{(k)}|\alpha_m|}{\sigma_w^2}\right)
}{
I_0\!\left(\tfrac{2y_m^{(k)}|r_m|}{\sigma_w^2}\right)
}
\Biggr].
\label{eq:LRT_loglike_int_exact}
\end{align}
This constitutes the optimal
quantum-limited detector for the RAQR under the magnitude-only sensing model.
Only the Bessel-ratio term depends on the observed data.  
The quadratic term
\(
C_{\text{GA}}
=\frac{K}{\sigma_w^2}
\sum_{m=1}^{N_r} (|r_m|^2-|\alpha_m|^2)
\)
is a deterministic offset for a fixed block.  
Moving it to the right-hand side yields the equivalent test
\begin{align}
T_{\mathrm{GA}}(\mathbf{Y}) = \sum_{m=1}^{N_r}\sum_{k=1}^{K}
\ln
\frac{
I_0\!\left(\tfrac{2 y_m^{(k)} |\alpha_m|}{\sigma_w^2}\right)
}{
I_0\!\left(\tfrac{2 y_m^{(k)} |r_m|}{\sigma_w^2}\right)
}
\underset{\mathcal{H}_0}{\overset{\mathcal{H}_1}{\gtrless}}
\tau^\star ,
\label{eq:GA_LRT_normalized}
\end{align}
with the normalized threshold
\begin{align}
\tau_{\text{GA}}^\star
= 
\ln\eta
-
\frac{K}{\sigma_w^2}
\sum_{m=1}^{N_r}\!\left(|r_m|^2 - |\alpha_m|^2\right).
\label{eq:GA_LRT_threshold}
\end{align}

This normalized form is used in simulations because it isolates all
measurement-dependent contributions of the RAQR (i.e., the quantum Rician
fluctuations) on the left, while absorbing the deterministic atomic–channel
offset into a single threshold~$\tau^\star$.

\subsection{Phase-Averaged LRT}

In practice, the transmit vector $\mathbf{x}$ belongs to the primary system and
is unknown to the RAQR-based sensing receiver.  
Under constant-modulus $M$-PSK signalling,
$x_\ell=\sqrt{P_\ell}\,e^{j\theta_\ell}$ with
$\theta_\ell$ uniformly drawn from the PSK alphabet,
the secondary receiver has no knowledge of the symbol phases.
Hence the noncentrality parameter
$|\alpha_m(\mathbf{x})| = |\mathbf{h}_m^T\mathbf{x}+r_m|$
in the genie-aided LRT \eqref{eq:LRT_loglike_int_exact} is unknown. 
A tractable approximation replaces $|\alpha_m(\mathbf{x})|$ with its
phase-averaged value
\(
\bar{\alpha}_m \triangleq 
\mathbb{E}_{\boldsymbol{\theta}}\!\left[\,|\mathbf{h}_m^T\mathbf{x}+r_m|\,\right].
\)
Writing $v_m=\sum_{\ell=1}^{N_t}h_{m,\ell}\sqrt{P_\ell}\,e^{j\theta_\ell}$, so
$\alpha_m=v_m+r_m$, the random phasor sum $v_m$ is well approximated as
$v_m\sim\mathcal{CN}(0,\sigma_{v,m}^2)$ with
\(
\sigma_{v,m}^2 = \frac{P}{N_t}\sum_{\ell=1}^{N_t}|h_{m,\ell}|^2,
\)
yielding the closed-form Rician mean $\bar{\alpha}_m
= \mathbb{E}[|\alpha_m|]$~\cite{Rice1948bell} as 
\[
\bar{\alpha}_m
=\sqrt{\frac{\pi\sigma_{v,m}^2}{4}}\,
L_{1/2}\!\left(-\frac{|r_m|^2}{\sigma_{v,m}^2}\right),
\]
where $L_{1/2}(\cdot)$ is a Laguerre polynomial.
Substituting $|\alpha_m|\!\to\!\bar{\alpha}_m$ in
\eqref{eq:LRT_loglike_int_exact} gives the implementable
phase-averaged LRT
\begin{align}
\ln\Lambda(\mathbf{Y})
\!=\!\!
\sum_{m=1}^{N_r}\sum_{k=1}^{K}
\Biggl[
\frac{|r_m|^2-\bar{\alpha}_m^{\,2}}{\sigma_w^2}
+
\ln\!\frac{
I_0\!\left(\tfrac{2y_m^{(k)}\bar{\alpha}_m}{\sigma_w^2}\right)
}{
I_0\!\left(\tfrac{2y_m^{(k)}|r_m|}{\sigma_w^2}\right)
}
\Biggr].
\label{eq:LRT_phaseavg}
\end{align}

As in the genie-aided case, the observation-independent term
\(
C_{\mathrm{PA}}
=\frac{K}{\sigma_w^2}
\sum_{m=1}^{N_r}\big(|r_m|^2-\bar{\alpha}_m^{\,2}\big)
\)
is a deterministic block offset.  
Moving it to the right-hand side yields the practical decision rule
\begin{align}
T_{\mathrm{PA}}(\mathbf{Y})
&=
\sum_{m=1}^{N_r}\sum_{k=1}^{K}
\ln
\frac{
I_0\!\left(\tfrac{2y_m^{(k)}\bar{\alpha}_m}{\sigma_w^2}\right)
}{
I_0\!\left(\tfrac{2y_m^{(k)}|r_m|}{\sigma_w^2}\right)
}
\underset{\mathcal{H}_0}{\overset{\mathcal{H}_1}{\gtrless}}
\tau_{\mathrm{PA}}^\star,
\label{eq:LRT_phaseavg_final}
\end{align}
with the shifted threshold
\begin{align}
\tau_{\mathrm{PA}}^\star
&= \ln\eta 
-
\frac{K}{\sigma_w^2}
\sum_{m=1}^{N_r}\!\left(|r_m|^2-\bar{\alpha}_m^{\,2}\right).
\label{eq:th_phaseavg}
\end{align}

The detector in \eqref{eq:LRT_phaseavg_final} depends only on
$(\mathbf{H},\mathbf{r},P)$ and thus provides a fully realizable
quantum-aware LRT for RAQR sensing.  
It preserves the functional structure of the optimal genie-aided test,
while eliminating dependence on the unknown—and rapidly varying—instantaneous
symbol phases of the primary transmitter.

\begin{remark}[Accuracy and Limitations of the Phase-Averaged LRT]
The phase-averaged approximation replaces the unknown noncentrality
$|\mathbf{h}_m^{T}\mathbf{x}+r_m|$ with its statistical mean and is therefore
implementable under realistic RAQR operation where the primary symbol phases
are unknown and vary rapidly.  
Because the Rician likelihood depends on the \emph{magnitude} of the complex
mean—not its power—the approximation is inevitably looser than the
genie-aided case, which uses the exact instantaneous value.  
Nevertheless, numerical results show that for multi-shot operation ($K\!\ge\!5$),
the phase-averaged LRT captures most of the achievable quantum-sensing gain,
providing a physically consistent detector for practical
RAQR-based spectrum monitoring.
\end{remark}

\section{Non-Coherent Energy Detector (ED)}\label{sec:ed}

In many quantum-sensing scenarios, a Rydberg-atom receiver cannot rely on
coherent MIMO information: the atomic ensemble does not provide access to the
RF carrier phase, the BS transmit vector $\mathbf{x}$ is unknown, and the
channel matrix $\mathbf{H}$ cannot be estimated from magnitude-only
measurements.  
This motivates a fully \emph{non-coherent} benchmark detector that uses only
the total optical energy collected across vapor cells and shots.

The decision rule for energy detector statistic is~\cite{atapattu2014energy}
\begin{equation}
T_{\mathrm{ED}}(\mathbf{Y})
= \sum_{m=1}^{N_r}\sum_{k=1}^{K} \bigl(y_m^{(k)}\bigr)^2
\underset{\mathcal{H}_0}{\overset{\mathcal{H}_1}{\gtrless}} \tau,
\label{eq:ED_test}
\end{equation}
which denotes the accumulated optical energy.  

\subsubsection{Distribution under $\mathcal{H}_0$ and $\mathcal{H}_1$}
Each magnitude measurement satisfies
\(
y_m^{(k)} = |\nu_{i,m} + w_m^{(k)}|\) with \(i\in\{0,1\},
\)
where $w_m^{(k)}\sim\mathcal{CN}(0,\sigma_w^2)$ and
\(
\nu_{0,m}=r_m,\;
\nu_{1,m}= \mathbf{h}_m^T\mathbf{x}+r_m.
\)
Thus,
\[
\frac{2}{\sigma_w^2}(y_m^{(k)})^2
\sim \chi^2_2(\lambda_{i,m}),\qquad
\lambda_{i,m}=\tfrac{2|\nu_{i,m}|^2}{\sigma_w^2}.
\]
Summing over $L=N_rK$ independent terms gives
\[
Z \triangleq \tfrac{2}{\sigma_w^2}T_{\mathrm{ED}}
\sim \chi_{2L}^2(\Lambda_i),
\qquad
\Lambda_i = K\sum_{m=1}^{N_r}\lambda_{i,m}.
\]

\subsubsection{Exact CFAR Threshold and Detection Probability}

Under $\mathcal{H}_0$, the test statistic
\(
Z \triangleq \tfrac{2}{\sigma_w^2} T_{\mathrm{ED}}
\)
follows a noncentral chi-square distribution
\(
Z \sim \chi_{2L}^2(\Lambda_0)
\)
with $L=N_rK$ degrees of freedom.  
Hence the false-alarm probability is
\begin{equation}
P_{\mathrm{FA}}\!=\! \Pr(Z > z_\tau \mid \mathcal{H}_0)
= Q_{L}\!\left(\sqrt{\Lambda_0},\,\sqrt{z_\tau}\right),
\,
z_\tau \!=\! \tfrac{2\tau}{\sigma_w^2},
\label{eq:ED_Pfa}
\end{equation}
where $Q_L(\cdot,\cdot)$ denotes the generalized Marcum--$Q$ function~\cite{nuttall1975some}. Since the secondary node lacks channel or waveform knowledge, it is
reasonable to operate the ED under a fixed CFAR level $P_{\mathrm{FA}}$.  
Solving \eqref{eq:ED_Pfa} for the threshold gives
\begin{equation}
\tau
= \frac{\sigma_w^2}{2}
\!\left[
Q_L^{-1}\!\Big(P_{\mathrm{FA}};\sqrt{\Lambda_0}\Big)
\right]^2 .
\label{eq:ED_tau}
\end{equation}

Under $\mathcal{H}_1$, the noncentrality parameter becomes $\Lambda_1$, yielding
\begin{equation}
P_{\mathrm{D}}
= \Pr(Z > z_\tau \mid \mathcal{H}_0)
= Q_{L}\!\left(\sqrt{\Lambda_1},\,\sqrt{z_\tau}\right).
\label{eq:ED_Pd_basic}
\end{equation}

Substituting the CFAR threshold \eqref{eq:ED_tau} into
\eqref{eq:ED_Pd_basic} produces the closed-form CFAR detection probability
\begin{equation}
P_{\mathrm{D}}(P_{\mathrm{FA}})
= 
Q_{L}\!\Big(
\sqrt{\Lambda_1},\;
Q_L^{-1}\!\big(P_{\mathrm{FA}};\sqrt{\Lambda_0}\big)
\Big).
\label{eq:ED_Pd_CFAR}
\end{equation}

\begin{remark}[Role and Limitations of the Non-Coherent ED]
The non-coherent ED uses only total optical energy and discards all
quantum–spatial structure of the RAQR.  
It does not exploit the coherent MIMO combining inherent in
$\mathbf{H}\mathbf{x}$, nor the nonlinear Rydberg response encoded in the
Rician likelihood.  
As such, it provides a deliberately conservative baseline for quantum sensing:
simple, physically robust, and implementable when the atomic receiver cannot
track phase or estimate the RF channel, but significantly suboptimal compared
with the LRT-based detectors that leverage RAQR physics.
\end{remark}

\subsubsection{Classical RF Energy Detector (Baseline for Comparison)}

To quantify the sensing advantage of Rydberg–atom receivers, we compare against the
\emph{classical RF energy detector}, which represents the standard non-coherent
baseline used in spectrum sensing, radar, and low-SNR emitter detection.
Unlike the RAQR, the RF receiver acquires coherent complex baseband samples,
has no reference-field injection, and operates under thermal-noise–limited
statistics. This contrast isolates the benefit of the RAQR’s quantum noise floor
and reference-assisted magnitude readout.

\subsubsection*{RF Observation Model}

The $m$-th RF sensor collects \(K_{\mathrm{RF}}\) complex baseband samples,
\[
y_{m}^{(k)}=
\begin{cases}
n_{m}^{(k)}, & \mathcal{H}_0,\\[1mm]
\mathbf{g}_m^{T}\mathbf{x} + n_{m}^{(k)}, & \mathcal{H}_1,
\end{cases}
\qquad k=1,\ldots,K_{\mathrm{RF}},
\label{eq:rf_multishot_model}
\]
where  
\(
n_m^{(k)} \sim \mathcal{CN}(0,\sigma_n^2)
\, \text{ i.i.d. across $m,k$},
\)
and \(\mathbf{g}_m\in\mathbb{C}^{N_t\times 1}\) is the RF channel to sensor \(m\),
distinct from the quantum-channel vector \(\mathbf{h}_m\) used by the RAQR.
The deterministic signal term under \(\mathcal{H}_1\) is
\(
\mu_m \triangleq \mathbf{g}_m^{T}\mathbf{x},
\)
where \(\mathbf{x}\) is the same (unknown) transmit vector seen by both
the RF and RAQR receivers.

\subsubsection*{Non-Coherent RF Energy Detector}

The RF energy detector forms the statistic
\begin{equation}
T_{\mathrm{ED}}^{(\mathrm{RF})}
= \sum_{m=1}^{N_r}\sum_{k=1}^{K_{\mathrm{RF}}} |y_m^{(k)}|^2
\underset{\mathcal{H}_0}{\overset{\mathcal{H}_1}{\gtrless}}
\tau_{\mathrm{RF}},
\label{eq:ED_test_RF}
\end{equation}
where $\tau_{\mathrm{RF}}$ is the threshold. 
Define the normalized test variable
\(
Z_{\mathrm{RF}}
\triangleq
\frac{2}{\sigma_n^2}\,T_{\mathrm{ED}}^{(\mathrm{RF})}\) with \(L = N_r K_{\mathrm{RF}}.
\)

\emph{Distribution under $\mathcal{H}_0$.}
When $y_m^{(k)}=n_m^{(k)}$, the statistic satisfies
\(
Z_{\mathrm{RF}} \mid \mathcal{H}_0 \sim \chi_{2L}^{2},
\)
so the false-alarm probability is
\begin{equation}
P_{\mathrm{FA}}
= Q_{L}\!\left(0,\,\sqrt{z_\tau}\right),
\qquad z_\tau = \tfrac{2\tau_{\mathrm{RF}}}{\sigma_n^2},
\label{eq:Pfa_RF}
\end{equation}
where $Q_L(\cdot,\cdot)$ is the generalized Marcum--$Q$ function.

\emph{Distribution under $\mathcal{H}_1$.}
Under $\mathcal{H}_1$, each sample has mean $\mu_m$, giving
\(
Z_{\mathrm{RF}} \mid \mathcal{H}_1 
\sim \chi_{2L}^{2}(\Lambda_{\mathrm{RF}}),
\)
with aggregate noncentrality parameter
\begin{equation}
\Lambda_{\mathrm{RF}}
= \sum_{m=1}^{N_r}
\frac{2K_{\mathrm{RF}}|\mu_m|^2}{\sigma_n^{2}}
=
\frac{2K_{\mathrm{RF}}}{\sigma_n^{2}}
\sum_{m=1}^{N_r}\!|\mathbf{g}_m^{T}\mathbf{x}|^2.
\label{eq:Lambda_RF}
\end{equation}

Hence the detection probability is
\begin{equation}
P_{\mathrm{D}}
= Q_{L}\!\left(\sqrt{\Lambda_{\mathrm{RF}}},\,\sqrt{z_\tau}\right),
\label{eq:Pd_RF}
\end{equation}
which serves as the classical baseline against which RAQR performance is evaluated.

\begin{remark}[RAQR vs.\ RF Energy Detection]
Classical RF receivers rely on coherent I/Q sampling and thermal-noise–limited
statistics, enabling arbitrarily large sample counts and asymptotic detection
gains. In contrast, a Rydberg-atom quantum receiver operates in the
quantum–optical regime: its measurements are magnitude-only, Rician distributed,
and constrained by atomic coherence times, allowing only a limited number of
independent shots. Despite this restriction, the RAQR benefits from a much lower
intrinsic noise floor and reference-assisted field readout, enabling reliable
weak-signal detection with far fewer samples than required by an RF front end.
Thus, the RF ED serves as a classical baseline, while the RAQR ED reveals the
fundamental sensing advantage arising from quantum-enabled field extraction.
\end{remark}

\section{Numerical Results}\label{sec_num}

We now evaluate the performance of the proposed multi-shot RAQR detectors. Unless otherwise stated, the system employs
$N_t=3$ transmit antennas and $N_r=4$ vapor-cell receivers with deterministic
$4$-PSK signalling. The carrier frequency is set to $28$\,GHz and is matched to
the Rydberg transition $62D_{5/2}\!\leftrightarrow\!64P_{3/2}$, whose
electric-dipole moment is
$\boldsymbol{\mu}_{eg}=[0,\ 789.1\,qa_0,\ 0]^\top$  
($q=1.602\times10^{-19}\,\mathrm{C}$,  
$a_0=5.292\times10^{-11}\,\mathrm{m}$).  
Quantum noise is modeled as
$w_m^{(k)}\!\sim\!\mathcal{CN}(0,\sigma_w^2)$, corresponding to a thermal-equivalent noise floor of
approximately $-97$\,dBm over a $\sim1$\,MHz optical detection bandwidth.
Channel coefficients $h_{m,\ell}\sim\mathcal{CN}(0,1)$ are normalized to
represent co-located Rydberg cells. The total transmit power $P$ is chosen so
that the effective sensing SNR is $0$\,dB based on the average Frobenius norm
of $\mathbf{H}$. 
All results are averaged over $10^5$ Monte Carlo realizations. We report either
the Bayesian total error probability
\(
P_e=\tfrac12(P_{\mathrm{FA}}+P_{\mathrm{MD}})
\)
with equal priors, or—under a CFAR design—the detection probability $P_D$ at
$P_{\mathrm{FA}}=0.1$. 



\begin{figure}[t!]
    \centering
    \includegraphics[width=0.8\linewidth]{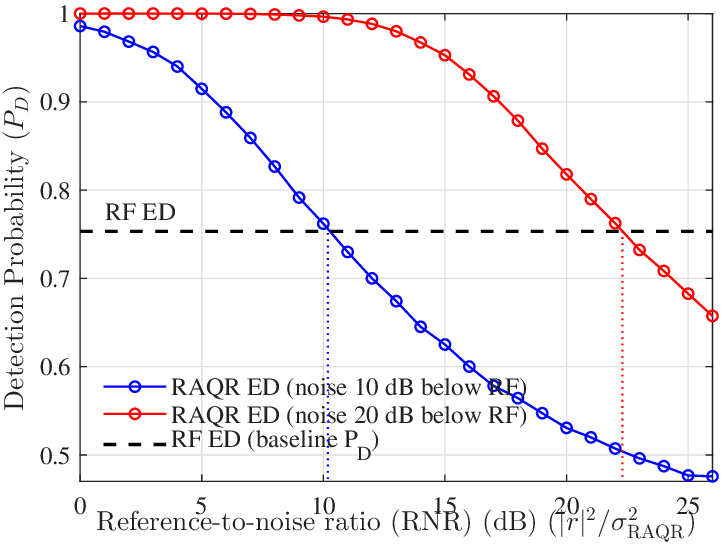}
    \caption{Detection probability $P_D$ versus reference--to--noise ratio (RNR) at CFAR $P_{FA}=0.1$ for RAQR and RF EDs.\vspace{-6mm}}
    \label{fig:PdvsRNR_ED}
\end{figure}

Figure~\ref{fig:PdvsRNR_ED} shows the detection probability $P_D$ at
$P_{FA}=0.1$ versus the reference--to--noise ratio
$\mathrm{RNR}=|r|^{2}/\sigma_{w,\mathrm{RAQR}}^{2}$.  
The RF energy detector exhibits a constant $P_D\!\approx\!0.75$ since its
central chi–square statistic is independent of any reference tone.  
In contrast, the RAQR statistic is noncentral (Rician), and its detectability
varies with the reference amplitude.  
With a $10$\,dB RAQR noise advantage, the RAQR ED outperforms the RF ED for
$\mathrm{RNR}\!\lesssim\!10$\,dB; with a $20$\,dB advantage, the crossover increases
to $\mathrm{RNR}\!\approx\!22$\,dB.  
These trends highlight that RAQR performance is maximized when the injected
reference is matched to its low intrinsic noise floor—overly strong references
suppress the signal contribution and degrade detection.

\begin{figure}[t!]
    \centering
    \includegraphics[width=0.80\linewidth]{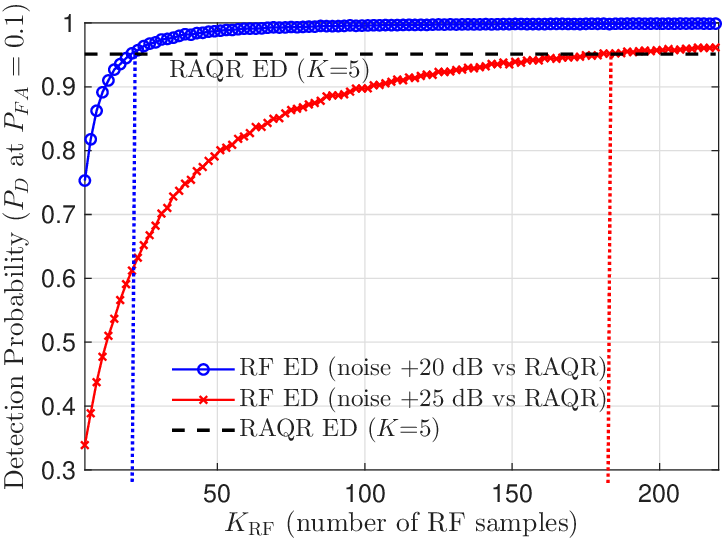}
    \caption{Detection probability $P_D$ versus number of RF samples $K_{\mathrm{RF}}$ at CFAR $P_{FA}=0.1$. 
    RAQR uses $K=5$. \vspace{-6mm}}
    \label{fig:Pd_vs_Krf_ED}
\end{figure}

Figure~\ref{fig:Pd_vs_Krf_ED} compares blind RAQR and RF energy detectors
under a CFAR constraint $P_{FA}=0.1$.  
With only $K_{\mathrm{RAQR}}=5$ shots, the RAQR achieves
$P_D\!\approx\!0.95$ due to its low intrinsic noise floor and reference-induced
noncentrality.  
An RF receiver with a $20$\,dB higher noise floor requires
$K_{\mathrm{RF}}\!\approx\!20$ samples to reach comparable performance, whereas a
$25$\,dB penalty pushes this requirement beyond $K_{\mathrm{RF}}\!\approx\!180$.
These results show that, in the blind setting, RAQR achieves high detection
probability with orders-of-magnitude fewer samples than a classical RF energy
detector subject to a higher noise.

\begin{figure*}[t!]
    \centering
    \subfloat[ROC for $K\!=\!1,5$.]{
        \label{fig:roc_multi_shot}
        \includegraphics[width=0.32\textwidth]{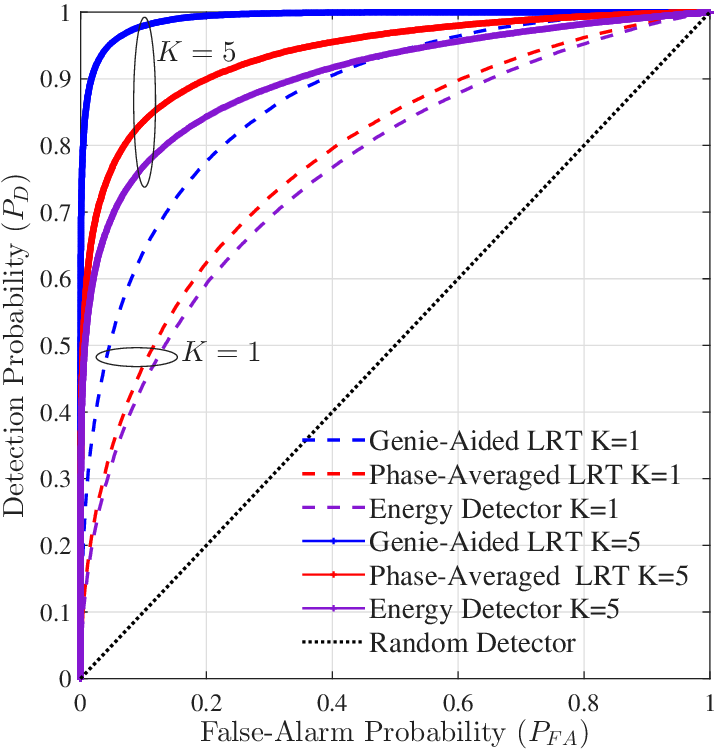}
    }
    \subfloat[$P_D$ vs.\ $K$ at CFAR $P_{\mathrm{FA}}=0.1$.]{
        \label{fig:pd_multi_shot}
        \includegraphics[width=0.32\textwidth]{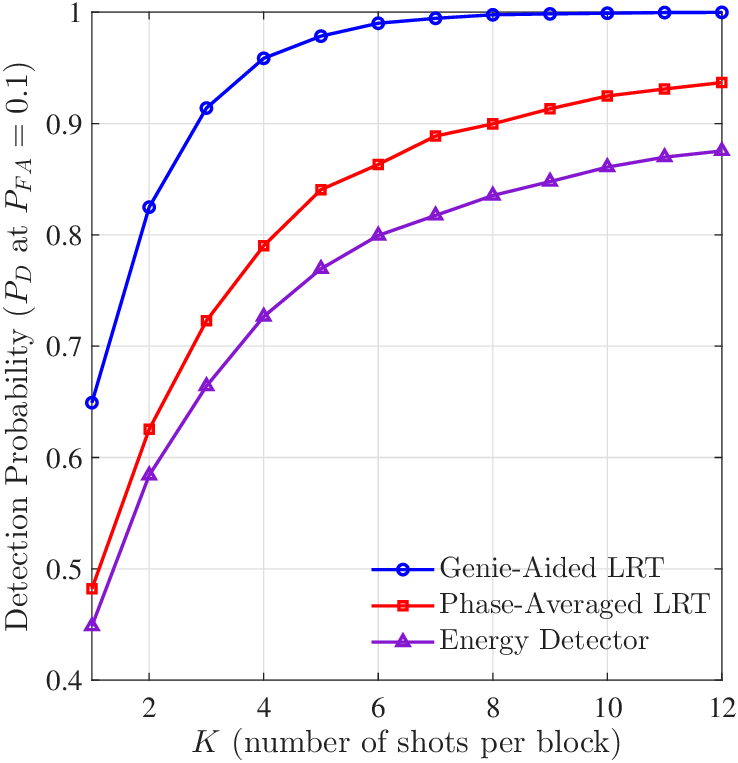}
    }
    \subfloat[$P_e$ vs.\ $K$ at $\eta=0.5$.]{
        \label{fig:Pe_vs_K_RAQR}
        \includegraphics[width=0.32\textwidth]{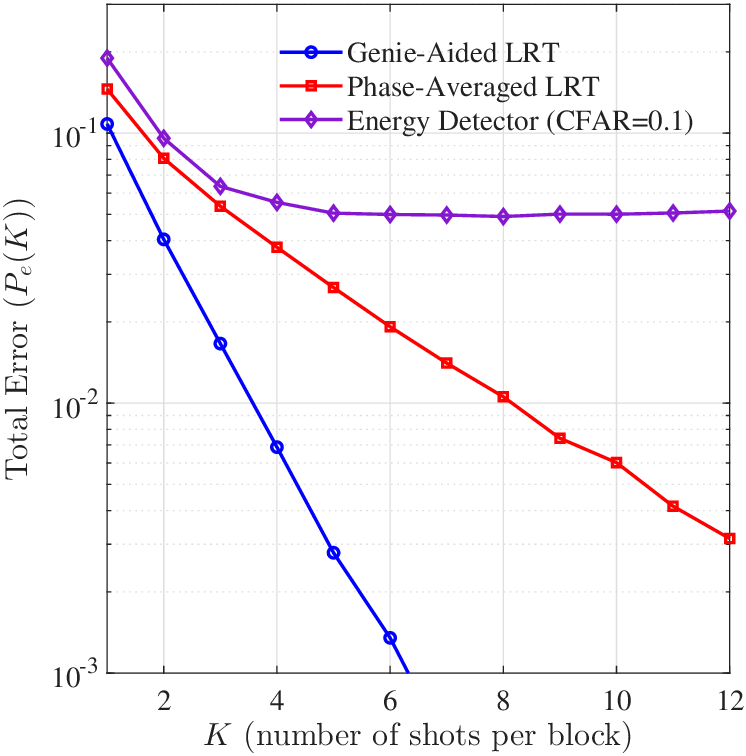}
    }
    \caption{Multi-shot RAQR detection performance.  
    (a) ROC improvement with shot aggregation;  
    (b) CFAR-constrained detection probability versus $K$;  
    (c) Bayesian total error versus $K$.  
    The demonstrate that few-shot averaging provides large practical gains. \vspace{-9mm}}
    \label{fig:RAQR}
\end{figure*}

Figure~\ref{fig:roc_multi_shot} compares the ROC performance of the RAQR
detectors for $K=1$ and $K=5$.  
The genie-aided LRT provides the fundamental bound; the phase-averaged LRT is
the practical coherent detector; and the blind ED serves as the non-coherent
baseline.  
At $P_{\mathrm{FA}}=0.1$, the phase-averaged LRT improves from
$P_D\!\approx\!0.5$ at $K=1$ to $P_D\!\approx\!0.85$ at $K=5$, a $\sim70\%$
relative gain.  
This ROC improvement highlights the role of multi-shot processing: single-shot
RAQR measurements are heavily noise-limited, while a small number of repeats
stabilizes the Rician statistics and markedly enhances hypothesis
separability.

Figure~\ref{fig:pd_multi_shot} plots $P_D$ versus the number of shots $K$ under
a CFAR constraint $P_{\mathrm{FA}}=0.1$.  
The genie–aided LRT provides the benchmark, reaching $P_D>0.95$ by $K\!\approx\!5$
and saturating near unity for $K\!\ge\!8$.  
The phase–averaged LRT increases steadily from $P_D\!\approx\!0.48$ at $K=1$ to
$P_D\!\approx\!0.94$ at $K=12$, closely approaching the genie bound as $K$
grows.  
The blind ED also improves with averaging but more slowly, rising from
$P_D\!\approx\!0.45$ at $K=1$ to $P_D\!\approx\!0.88$ at $K=12$.  
These results show that the practical phase–averaged LRT captures most of the
genie–aided performance with only a modest number of quantum shots, whereas
the ED remains fundamentally constrained by its noncoherent nature.

Figure~\ref{fig:Pe_vs_K_RAQR} shows the Bayesian total error versus the number of
shots~$K$.  
At $K=1$, the genie, phase-averaged, and ED detectors achieve
$P_e\!\approx\!0.11$, $0.15$, and $0.19$, respectively.  
As $K$ increases, the genie error drops fastest ($P_e\!\approx\!2.8\times10^{-3}$
at $K=5$ and $10^{-4}$ at $K=10$), while the phase-averaged LRT decreases to
$P_e\!\approx\!2.7\times10^{-2}$ at $K=5$ and $6\times10^{-3}$ at $K=10$.  
The ED improves only modestly and is limited by its noncoherent CFAR floor
around $0.05$.  
Overall, multi-shot processing is crucial for RAQR detection, and the
phase-averaged LRT captures most of the genie gain while remaining fully
implementable.

\begin{remark}[Limits of Multi--Shot RAQR Acquisition]\label{re_limit}
Unlike RF receivers that can collect hundreds of independent samples per
coherence interval, RAQRs support only a small number of usable shots
($K\!\lesssim\!10$--$20$) before atomic backaction and collisional dynamics
introduce temporal correlation. Beyond this regime, the i.i.d.\ Rician model
breaks down and the likelihood becomes a correlated, non-Gaussian of
magnitude-only measurements. Developing tractable detectors and statistical
models for this correlated quantum readout regime remains an important open
direction for future RAQR sensing theory.
\end{remark}

\section{Conclusion}\label{sec:conclusion}

This paper presented a unified statistical framework for multi–shot detection with
Rydberg–atom quantum receivers.  
Using the magnitude–only optical readout imposed by atomic EIT, we derived the exact
multi–shot LRT, a practical phase-averaged LRT that requires no knowledge of the
unknown RF waveform, and a fully non-coherent energy detector with closed-form CFAR
characterization.  
The analysis and simulations show that only a handful of quantum shots are sufficient
to stabilise the Rician measurement statistics, allowing the implementable
phase-averaged LRT to approach genie-aided performance and substantially outperform
classical RF energy detectors operating under realistic noise figures.  
These results demonstrate that multi-shot quantum processing is essential for
unlocking the sensing capability of RAQRs and establish a rigorous bridge between
quantum measurement constraints and classical detection theory. 
Future directions include modelling back-action–limited shot regimes, incorporating
temporal noise correlations, extending the framework to wideband and multi-band RF
fields, and exploring adaptive reference-field design.  
Such developments will further clarify the role of quantum receivers as front-end
components in next-generation quantum–enhanced sensing for
low-SNR RF systems.



\end{document}